\documentclass[useAMS,usenatbib]{mn2e}
\usepackage{amsmath}
\usepackage{graphicx}

\newcommand{\ms}{\rm {h^{-1}M_{\odot}}}
\newcommand{\mpch}{{\rm h^{-1}Mpc}}

\begin{document}
\bibliographystyle{mn2e}
\graphicspath{{./fig/}}
\title[]{First Galaxy-Galaxy Lensing Measurement of Satellite Halo Mass in the CFHT  Stripe-82 Survey}
\author[Ran Li et. al]
       {\parbox[t]{\textwidth}{
        Ran Li$^{1}$\thanks{E-mail:ranl@bao.ac.cn},
        Huanyuan Shan$^{2}$, 
        Houjun Mo$^{3}$,
        Jean-Paul Kneib$^{2,4}$,
        Xiaohu Yang$^{5,6}$,
        Wentao Luo$^{6}$,
        Frank C. van den Bosch$^{7}$,
        Thomas Erben$^{8}$,
        Bruno Moraes$^{9}$,
        Martin Makler$^{9}$}
        \vspace*{3pt} \\
  $^{1}$National Astronomical Observatories, Chinese Academy of
           Sciences, 20A Datun Rd, Chaoyang District, Beijing, China 100012\\
  $^{2}$Laboratoire d'Astrophysique, Ecole Polytechnique Fédérale de Lausanne (EPFL), 
          Observatoire de Sauverny, CH-1290 Versoix, Switzerland\\
  $^{3}$Department of Astronomy,  University of Massachusetts, Amherst MA 01003, USA \\
  $^{4}$Aix Marseille Université, CNRS, LAM (Laboratoire
          d'Astrophysique de Marseille) UMR 7326, 13388, Marseille, France\\
  $^{5}$Center for Astronomy and Astrophysics, Shanghai Jiao Tong University, Shanghai 200240, China\\
  $^{6}$Key Laboratory for Research in Galaxies and Cosmology, Shanghai Astronomical Observatory,
            Nandan Road 80, Shanghai 200030, China\\
  $^{7}$Astronomy Department, Yale University, PO Box 208101, New Haven, CT 06520-8101, USA\\
  $^{8}$Argelander Institute for Astronomy - University of Bonn, Auf dem H\"ugel 71, 53121 Bonn Germany\\
  $^{9}$Centro Brasileiro de Pesquisas F\'{i}sicas, Rua Dr. Xavier Sigaud 150, CEP 22290-180, Rio de Janeiro, RJ, Brazil}

\maketitle

\begin{abstract}
  We select satellite galaxies from the galaxy group catalog
  constructed with the SDSS spectroscopic galaxies and measure the
  tangential shear around these galaxies with the source catalog extracted
  from the CFHT Stripe-82 Survey. Using the tangential shear, 
  we  constrain the mass of subhalos  associated with these satellites.  
  The lensing signal is measured around satellites in groups 
  with masses in the range $10^{13}$ - $5 \times 10^{14}\,\ms$, 
  and is found to agree well with theoretical expectations.  Fitting
  the data with a truncated NFW profile, we obtain an average 
  subhalo mass of $\log (M_{\rm sub}/\ms) = 11.68\pm 0.67$ for 
  satellites whose projected distances to central
  galaxies are in the range $0.1$ - $0.3\,\mpch$, 
  and $\log (M_{\rm sub}/\ms) = 11.68\pm 0.76$ for satellites with projected
  halo-centric distance in $[0.3, 0.5]\,\mpch$.  The best-fit subhalo
  masses are comparable to the truncated subhalo masses assigned to
  satellite galaxies using abundance matching and about 5 to 10 times
  higher than the average stellar mass of the lensing satellite
  galaxies.
\end{abstract}


\section{Introduction}

According to the cold dark matter (CDM) paradigm of structure
formation, dark matter halos form hierarchically through merging and
accretion, while galaxies form at the centers of dark matter halos
through gas accretion and star formation. When a small halo merges
into a larger one in such a hierarchical formation process, it becomes
a subhalo and may suffer environmental effects from the host, such as
tidal stripping and impulsive heating, that tend to disrupt
it. However, some subhalos may survive such processes and exist as the
halos of satellite galaxies at the present time. Investigations of the
masses and density profiles of subhalos can, therefore, provide
important test for the CDM scenario of structure formation.

The difficulty of measuring the dark matter distribution around 
galaxies arises from the dearth of proper tracers. For nearby field 
galaxies,  dynamical tracers such as satellites or HI clouds can 
be used to probe the host dark matter density profile
\citep[e.g.][]{Sofue2001,Buote2002,Prada2003}.
For subhalos, however, the observation is more difficult due 
to their relative low masses in comparison to their host halos. 
Gravitational lensing, which is sensitaive to surface mass 
density gradiant, may provide a promising way to study 
dark matter subhalos in their host halos. The existence of substructures
(e.g. subhalos) can produce flux-ratio anomalies in multiple images in
strong gravitational lensing systems
\citep{Mao1998,Metcalf2001,Mao2004,Kochanek2004,Maccio2006,xu2009},
can perturb the locations, and change the multiplicity of the lensed
images \citep{kneib1996,kneib2011}, and can disturb the surface
brightness of extended arcs and Einstein
rings \citep{Koopmans2005,Vegetti2009a,Vegetti2009b,Vegetti2010,Vegetti2012}.
Unfortunately the number of high quality images of strong lensing
systems is still limited, and strong lensing effects can only probe
the central regions of dark matter haloes\citep{kneib2011}. 
Consequently, quantitative constraint on subhalo properties 
has yet to be obtained from strong gravitational lensing observations.

Since subhalos are expected to be associated with satellite galaxies,
an alternative approach is to study the subhalo population in a
statistical way using galaxy-galaxy weak lensing
\citep{Yang2006,Limousin2007, Limousin2009, Natarajan2007,Li2009,
  Natarajan2009,Mira2011,Li2013,Gillis2013a}. With the advent of wide
and deep galaxy surveys, such as the Sloan Digital Sky Survey
(SDSS)\footnote{http://www.sdss.org} and the Canada-France-Hawaii
Telescope Legacy Survey
(CFHTLS)\footnote{http://www.cfht.hawaii.edu/Science/CFHTLS/},
galaxy-galaxy lensing can now be used to study the mass distribution
around lens galaxies of different luminosities, stellar masses,
colors, and morphological types \citep[e.g.,][]{Brainerd1996,
Hudson1998,Mckay2001,Hoekstra2003,Hoekstra2004,Mandelbaum2005,
Mandelbaum2006,Mandelbaum2008,Sheldon2009, Jonhston2007,Leauthaud2012}. 
However, even within a narrow luminosity and morphology range, 
a galaxy can either be a central galaxy located near the center 
of a dark matter halo, or a satellite galaxy associated with a 
dark matter subhalo. Thus, such galaxy-galaxy
lensing results do not measure directly the lensing signals of
subhalos alone, but rather the total signals produced by a mixture of
central and satellite galaxies \citep[e.g.][]{George2012,Gillis2013b}.
 
In \citet[][hereafter L13]{Li2013}, we have proposed a method to
measure the galaxy-galaxy lensing effect of subhalos by using
satellite galaxies selected from galaxy groups identified from the
SDSS spectroscopic catalog \citep[][]{Yang2005,Yang2007}. With such a
group catalog, one can not only distinguish satellites from centrals,
but also select lensing satellite galaxies both according to their
host halo masses and their projected distances to the host halo
center.  In this paper, we apply the method of L13 to real lensing
data obtained from the CFHT Stripe-82 Survey (CS82)
\citep[see][]{Comparat2013} together with the SDSS group
catalog of \citep{Yang2007}. To ensure a significant detection with
the current limited data, we select satellite galaxies from relatively
massive groups.

The paper is organized as follows. In \S\ref{sec:data}, we describe
the lens selection, the source catalog  and show the observation result.
In \S\ref{sec:model}, we present our theoretical model. 
In \S\ref{sec:result} we  compare the observation data
with model predictions to estimate the subhalo mass.  Finally, we
summarize our main results in \S\ref{sec:sum}.  Throughout the paper,
we adopt a $\Lambda$CDM cosmology with parameters given by the
WMAP-7-year data \citep{wmap7}.

\section{Observational Data}
\label{sec:data}

\subsection{The Source Catalog}
  
The Canada French Hawaii Telescope (CFHT)
Stripe-82 Survey is an $i$-band survey, 
which covers the SDSS equatorial Stripe82 region, and 
has a depth of $i_{AB}=24.0$ with excellent seeing conditions 
(between $0.4$ and $0.8$~arcsec with a median of $0.59$~arcsec). 
The survey, referred to as CS82 in the following, contains a 
total of $173$ tiles ($165$ tiles CFHT/Stripe82 and $8$ CFHT-LS Wide tiles). 
Each CS82 tile was obtained in four dithered observations 
with an exposure time of $410$s, each resulting in a 
$5$-$\sigma$ limiting magnitude in about 2~arcsec 
diameter aperture of about $i_{AB}=24.0$. After masking 
out bright saturated stars and other artifacts across 
the entire survey, the final
effective sky coverage is $\sim 124~{\rm deg}^2$.

The shape of source galaxies are measured with LENSFIT 
method \citep{Miller2007, Miller2013}, the details of the 
calibration and systematics of which are shown and 
discussed in \citet[][]{Heymans2012}. The data processing  
closely follows the procedures outlines in \citet{ Erben2009, Erben2013}. 
Specific procedures applied to the CS82 imaging will be described in 
Erben et al. (2014, in preparation).

In our work, the source galaxies are selected  with magnitudes 
$i_{AB}<23.5$, signal-to-noise $\nu>10$, weight $w>0$ and 
FITCLASS$=0$, where $w$ represents the inverse variance weight 
accounting for the intrinsic ellipticity distribution of the source 
galaxies and  FITCLASS is the object classification provided by LENSFIT. 
We obtain the photometric redshifts for our source galaxies from 
overlapping multi-color data of Sloan Digital Sky Survey. 
We further remove source galaxies with photometric redshift 
$z < 0.25$ to reduce the systematics brought by catastrophic outliers.
These criteria result in a total of $2,052,507$ source galaxies.

\subsection{Lens Selection}

To select galaxies according to their positions in halos, we use 
the group catalog constructed by \citet[][hereafter Y07]{Yang2007} 
from the SDSS Data Release 7 \citep{Abazajian2009} (hereafter SDSSGC
\footnote{http://gax.shao.ac.cn/data/Group.html}).
The group catalog is constructed with the adaptive halo-based group 
finder developed by \citet{Yang2005,Yang2007} using galaxies with 
spectroscopic redshifts in the range of $0.02 \le z \le 0.2$, and  
with redshift completeness $\mathcal{C} > 0.7$. Three group samples 
with different sources of galaxy redshifts have been constructed.  
Our analysis is based on Sample II which is based on all galaxies 
with spectroscopic redshifts either from the SDSS or from other 
sources. There are in total 18,217 galaxies in the CS82 
region, and a total of 13,978 groups including those 
with only one member.\footnote{Following Y07, we refer to a system 
of galaxies as a group regardless of its richness and mass, including 
isolated galaxies (i.e., groups with one member) and clusters 
of galaxies.}

Each of the groups in the SDSSGC has an assigned halo mass, $M$,
given by the ranking of its characteristic stellar mass, 
$M_{\rm stellar}$, defined to be the total stellar mass of member 
galaxies with $^{0.1}{\rm M_r} - 5 \log{\rm h} \le -19.5$, 
where $\rm ^{0.1}M_r$ is the absolute $r$-band magnitude 
with K-correction and evolution-correction to $z = 0.1$.  
The stellar mass of an individual galaxy is calculated 
with its magnitude and colors using the fitting formula 
given by \citet{Bell2003}. We refer readers to 
\citet{Yang2007,Yang2008} for the details of the group catalog 
and the halo mass assignment. 

For each group, the central galaxy is defined to be either the one
with the largest stellar mass or the one with the largest luminosity.
In our sample, we only use groups for which the brightest galaxies are
the same as the most massive galaxies. This criteria reduces our group
number by $10\%$. Galaxies other than centrals are called satellites.
We select satellite galaxies in groups with assigned masses in the
range $10^{13}$ - $5 \times 10^{14}$. We bin satellite galaxies
according to their projected halo-centric radii $r_p$, and the number
of satellites in each bin is listed in Table~\ref{tab:para}.

\subsection{Lensing Signal Computation}

In the weak lensing regime, the tangential shear, $\gamma_t(R)$, 
is related to the excess surface mass density, $\Delta\Sigma$, through
\begin{equation}\label{eq:ggl}
\Delta\Sigma(R)=\gamma_t(R)\Sigma_{\rm crit}={\overline\Sigma}(<R)-\Sigma(R)\,,
\end{equation}
where ${\overline\Sigma}(<R)$ is the average surface mass density
within $R$, and $\Sigma(R)$ is the average surface density at $R$. The
critical surface density can be written in terms of comoving
coordinates as
\begin{equation}
\Sigma_{\rm crit}=\frac{c^2}{4\pi G}\frac{D_s}{D_l D_{ls}(1+z_l)^2}\,,
\end{equation}
where $z_l$ is the redshift of the lens, $D_{ls}$ is the angular
diameter distance between the source and the lens, and $D_l$ and $D_s$
are the angular diameter distances from the observer to the lens and
to the source, respectively. The factor $(1 + z_l)^{−2}$  is due to the
use of comoving coordinates. To obtain
$\Delta\Sigma$ we stack lens-source pairs in 16 logarithmic radial
($R$) bins from $0.05$ to $2 \mpch$.  Only sources with photometric
redshifts $z_{s}-z_{l} > 0.1$ are used for a lens with redshift $z_l$.
For a sample of selected lenses, $\Delta\Sigma(R)$ is estimated using
\begin{equation}
\Delta\Sigma(R)=\frac{\sum_{ls}w_{ls}\gamma_t^{ls}\Sigma_{\rm crit}}{\sum_{ls}w_{ls}}\,,
\end{equation}
where
\begin{equation}
w_{ls}=w_n\Sigma_{\rm crit}^{-2}\,,
\end{equation}
with $w_n$ a weight factor, defined by Eq.\,(8) in
\citet{Miller2013} and introduced to account for  intrinsic
scatter in ellipticity and shape measurement error.
 
\subsection{Observational Results}

Fig.~\ref{fig:ESD1} shows the lensing signal around satellite galaxies
located in two different bins of $r_p$ in groups of masses $[10^{13},
5\times10^{14}] \ms$, where $r_p$ is the projected distance between
the satellite galaxy and the halo center. Although the error bars are large at small $R$,
the two components in the lensing signal as shown in figure 2 of L13
is evident here.  The satellite contribution dominates the
central part, decreases to a minimum when $R$ is about the average
$r_p$.  The lensing signal then rises when the host halo mass profile
starts to take over.  Errorbars shown are $1 \sigma$ fluctuations
obtained with a bootstrap method.  For clarity, the data points are
re-binned in $R$. The solid lines, which show a similar behavior 
as the data, are theoretical predictions which we discuss in detail below.

For reference we also show the lensing signal around the central
galaxies in these groups, i.e. those with assigned halo masses in the
range $[10^{13}, 5\times10^{14}] \,\ms$, and the result is shown in
Fig.~\ref{fig:central}.   The solid line shows the NFW halo
  \citep{NFW97}, with a concentration parameter given by the model of
  \citet{Neto2007}, that best fits the observational data. The
  corresponding halo mass is $\log(M/\ms) = 13.32$, which is in
  excellent agreement with the average of the assigned halo masses of
  the groups, which is $\log(M/\ms) = 13.37$.

\section{The Lens Model}
\label{sec:model}

We use the same method as in L13 to model the lensing signal around
satellite galaxies. Here we give a brief description of the method.

The mean tangential shear around a sample of galaxies is determined by
the average surface density $\Sigma(R)$, which is related to average
density profile, $\rho_{\rm g,m}$, around the galaxies.  Under the
approximation that the distances between the lenses and the 
observer are much larger than $R$, we can write:
\begin{equation}\label{xi_gm}
\Sigma(R)=\int\rho_{\rm g,m} \left(\sqrt{R^2+\chi^2} \right)\,{\rm d}\chi\,;
\end{equation}
and
\begin{equation}\label{xi_gm_in}
\Sigma(< R)=\frac{2}{R^2} \int_0^{R} \Sigma(u) \, u \, {\rm d}u ,
\end{equation}
where $\chi$ is the comoving distance along the line of sight.  The
 excess surface density,  $\Delta\Sigma$, at a distance $\rm R$
from a satellite galaxy can be written as:
\begin{equation}
\Delta\Sigma(R)=\Delta\Sigma_{\rm sub}(R)+\Delta\Sigma_{\rm host}(R, r_p)+\Delta\Sigma_{\rm star}(R)\,,
\end{equation}
where $\Delta\Sigma_{\rm sub}(R)$ is the contribution of the
  subhalo associated with the satellite, 
$\Delta\Sigma_{\rm host}(R, r_p)$ is the contribution from the host halo, where $r_p$ is the
projected distance between the satellite galaxy and the center of
the host halo,  and $\Delta\Sigma_{\rm star}(R)$ is the contribution of 
stellar component of the satellite.  For central galaxies, $\Delta\Sigma_{\rm sub}(R)$
vanishes.  We neglect the two-halo term, i.e. the contribution to the
lensing signal from other halos in the foreground and background. Our
previous studies \citep{Li2009,Cacciato2009} have shown that the
two-halo term is completely negligible on the scales we are concerned
with here.

For each satellite galaxy, the group halo mass is obtained from the
group catalog. We assume that the host halo is centered on the central
galaxy and has a NFW profile \citep{NFW97} with a concentration
parameter given by \citet{Neto2007}.  We use the subhalo mass function
given in \citet{Bosch2005} to assign mass to each satellite halo using
the abundance matching method described in L13. The subhalo density
profile is modeled with a truncated NFW profile,
\begin{equation}\label{eq:rhosub} 
\rho_{\rm sub}(r)=\left\{
\begin{array}{rl} f_t \, \rho_{i,{\rm sub}}(r|m_i) & r \le r_t,\\
    0 &  r > r_t
\end{array} \right. 
\end{equation}
where $\rho_{i,{\rm sub}}(r)$ is the NFW profile of the subhalo at the
time of accretion, and $m_i = M_{\rm sub}/f_m$ is the subhalo mass at
accretion. Following \citet{gao2004, Yang2006}, the parameter $f_m$, which
describes the retained mass fraction of the subhalo, is given by
\begin{equation}
f_m = 0.65 \, (r_{\rm dis}/r_{\rm vir})^{2/3}\,,
\end{equation}
where $r_{\rm dis}$ is the 3-D halo-centric distance, 
and $r_{\rm vir}$ is the virial radius of the host halo.   In the case of
real data, only the projected halo-centric distance can be
obtained. In order to obtain $r_{\rm dis}$, for a satellite with given
$r_p$ we randomly sample a 3-D halo-centric distance assuming that the
spatial distribution of satellites follows the NFW form.  The
parameter $f_t$ in equation (\ref{eq:rhosub}) describes the reduction
in the central density of the subhalo, and $r_{t}$ is the truncation
radius due to the tidal force of the host halo. The original density
profile of the subhalo at the time of accretion, 
$\rho_{i,{\rm sub}}(r)$, assumed to have a NFW form, is characterized by a scale
radius, $r_{\rm s,sub}$, and a characteristic density, 
$\delta_{0,{\rm sub}}$. Note that the parameters $f_t$ and $\delta_{0,{\rm sub}}$
can be combined into a single parameter, $\rho_{\rm 0,{\rm sub}}$. For
the truncation radius, $r_{t}$, we use the analytical tidal radius
formula
\begin{equation}\label{eq:rt}
r_t=\left( \frac{M_{\rm sub}}
{(2 - {\rm d}\ln M/{\rm d}\ln r)M(<r_{\rm dis})}  \right)^{1/3} r_{\rm dis}\,,
\end{equation}
where $M(<r_{\rm dis})$ is the host halo mass within a sphere of
radius $r_{\rm dis}$ \citep{BT87,Tormen98}. As shown by 
\citet{springel08_aqu}, this analytical model agrees well with the 
truncation radii of dark matter subhalos in $N$-body simulations.  
The density profile is normalized to the mass assigned to the subhalo 
by choosing a proper $f_t$ (or equivalently $\rho_{\rm 0,sub}$). 

Therefore, the satellite halo profile is specified by three
parameters: (i) the stellar mass of the satellite; (ii) the host halo
mass; and (iii) the projected halo-centric distance. For each
individual satellite in the group catalog, we can then calculate its
lensing signal with the model given above.  Averaging the signal for
selected satellite samples, we can make theoretical predictions which
can be compared to the observational signal to determine model
parameters.

Since the smallest scale probed in this work is $\sim 50$ kpc, 
much larger than the typical size of a galaxy,  we model the lensing 
signal from the stellar content of the satellite as that from a point 
source for simplicity. We can then write:
\begin{equation}
\Delta\Sigma_{\rm star}= \langle M_{\rm star} \rangle / R^2 \,,
\end{equation}
where $M_{\rm star}$ is the stellar mass of the satellite galaxy, and 
the angle bracket represents the average over the sample of satellites.

\begin{figure}
\includegraphics[width=0.5\textwidth]{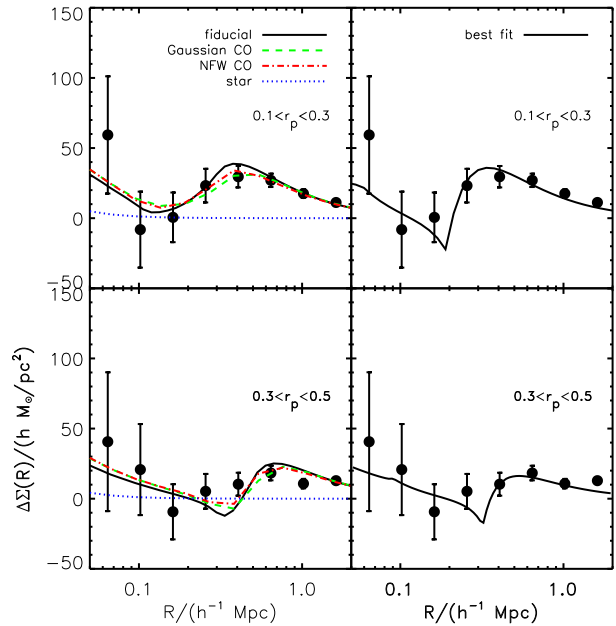}
\caption{Solid dots show the measured excess surface density profiles,
  $\Delta\Sigma(R)$, around satellite galaxies in groups with assigned
  masses in the range $[10^{13}, 5 \times 10^{14}]\, \ms$.  In the
  upper panels, the $r_p$ range of satellite galaxies is
  $[0.1,0.3]\mpch$, while in the lower panels, the range is
  $[0.3,0.5]\mpch$. Errorbars reflect the $68\%$ confidence intervals
  obtained using bootstrapping. The black solid lines in the left-hand
  panels show the predictions of our fiducial model (see
  \S\ref{sec:model}), while the dashed and dash-dotted lines show the
  predictions using two different models for the center-offsets (see
  text for details). The blue dotted lines show the contribution of stellar component.
  The right-hand panels show the same data, but
  this time the solid lines correspond to the best-fit models
  described by Eq.(\ref{eq:fit}). The corresponding best-fit
  parameters are listed in Table~\ref{tab:para}.}
\label{fig:ESD1}
\end{figure}

\begin{figure}
\includegraphics[width=0.5\textwidth]{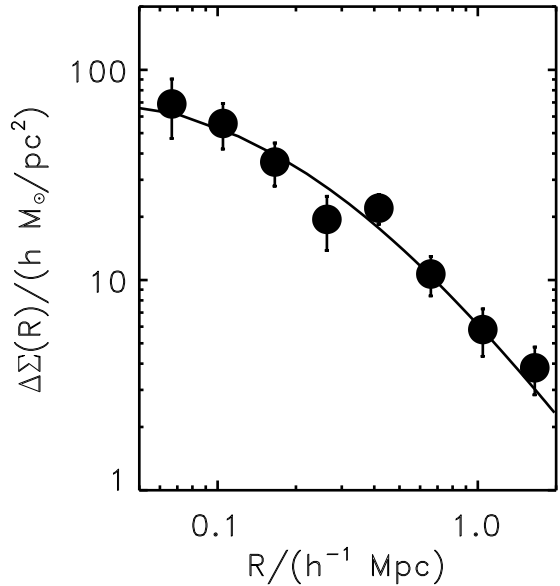}
\caption{Solid dots show the measured $\Delta\Sigma(R)$ around central
  galaxies in groups with assigned masses in the range $[10^{13}, 5
  \times 10^{14}] \,\ms$.  Errorbars reflect the 1$\sigma$
  uncertainties. The solid line corresponds to the excess surface
  density of the best-fit single NFW profile, which has a mass
  $\log(M/\ms) = 13.32$, in excellent agreement with the average of
  the assigned halo masses of the groups, which is $\log(M/\ms) =
  13.37$.}
\label{fig:central}
\end{figure}

The model predictions thus obtained are shown as solid lines in the
left-hand panels of Fig.~\ref{fig:ESD1}. Note that the model is in
good agreement with the data, even without any fitting.  The reduced $\chi^2$
values for  $[0.1,0.3]$ and $[0.3,0.5]\,\mpch$ $r_p$ bins are 1.3 and 1.8 respectively.
The blue dotted line represents the contribution of the stellar components, which
accounts for $\sim 10\%$ of the predicted lensing signal at the inner region. 
This fiducial model assumes that all central galaxies reside exactly at the
centers of their dark matter host haloes. 
  
However, various studies
have shown that in reality, central galaxies can often be offset
from the center of their host halo
\citep[e.g.][]{vdBosch2005,Skibba2011,George2012}.  In particular,
the recent study by \cite{Wang2013} found that $\sim 20\%$ of
central galaxies are offset from the center of their dark matter
halo, and that the offsets roughly follow a NFW profile with a
concentration parameter $c \sim 6$. To test the potential impact of
such center-offsets on the lensing signal studied here, we consider
two different models: (1) we assume that all
central galaxies have an offset, $\Delta r$ that follows a Gaussian
distribution centered on $\Delta r = 0$, and with a standard
deviation of $0.1\, \mpch$; (2) we assume that only $20\%$ of the central
galaxies have a non-zero offset, and that the probability
distribution for their $\Delta r$ follows a NFW profile with
concentration parameter $c=6$. The dashed and dotted lines in the
left-hand panels of Fig.~\ref{fig:ESD1} show the model predictions
for offset models (1) and (2), respectively.  Note how
the center-offsets `smooth' out the contribution of the host halo to the
overall lensing signal.  Both models yield results that are
virtually indistinguishable, and in even better agreement with the
data than our fiducial model without center-offsets.


\section{Constraints on Model Parameters}
\label{sec:result}

\begin{figure}
\includegraphics[width=0.5\textwidth]{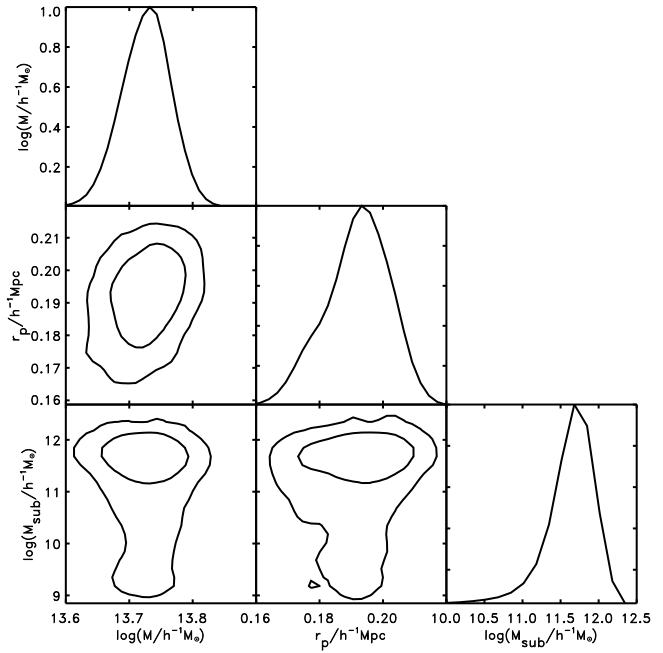}
\caption{The 68\% and 95\% confidence intervals for parameters $M$,
  $r_p$ and $M_{\rm sub}$. The last panel in each row shows the
  marginalized posterior distribution. The $r_p$ range of satellite
  galaxies is $[0.1,0.3]\mpch$.}
\label{fig:mcmc1}
\end{figure}

\begin{figure}
\includegraphics[width=0.5\textwidth]{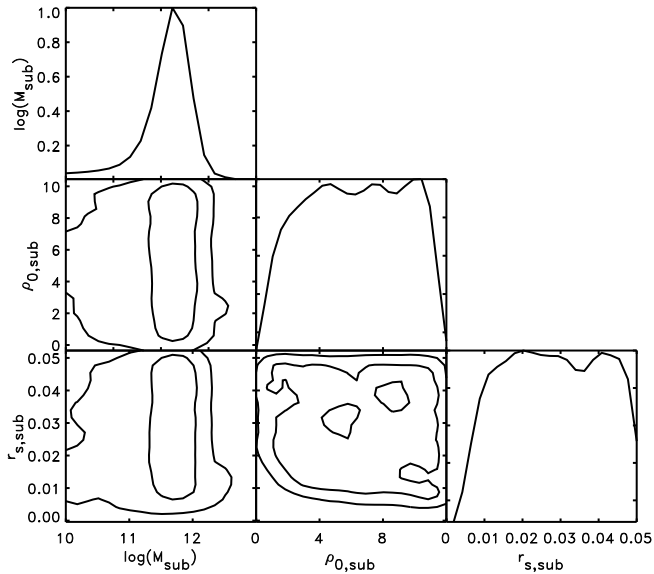}
\caption{Similar to Fig.~\ref{fig:mcmc1}, but here the 68\% and 95\%
  confidence intervals are shown for parameters $M_{\rm sub}$,
  $\rho_{\rm 0,sub}$ and $r_{\rm s,sub}$. }
\label{fig:mcmc2}
\end{figure}

Our theoretical predication given above is obtained by modeling halos
and subhalos of individual satellites.  In order to see how the
observational data constrain the average mass of subhalos, we fit the
data with a simple model assuming that the average lensing signal has
the same  form as a single lens system:
\begin{equation}
\label{eq:fit}
\Delta\Sigma=\Delta\Sigma_{\rm host}(R|M,r_p) + 
\Delta\Sigma_{\rm sub}(R|M_{\rm sub}, \rho_{\rm 0,sub}, r_{\rm s,sub})\,.
\end{equation} 
Since the lensing signal of the stellar component is much smaller than that
from dark matter subhalo, we ignore this component for simplicity. 
The model is therefore described by 5 free parameters: the host halo
mass, $M$; the projected halo-centric distance $r_p$; the subhalo mass
$M_{\rm sub}$; the subhalo characteristic density $\rho_{\rm 0,sub}$;
and the subhalo scale radius $r_{\rm s,sub}$. For simplicity, the
concentration of the host halo is fixed using the concentration-mass
relation of \citet{Neto2007}.  We also ignore the center-offset here 
because it  does not affect the lensing signal significantly.
As in L13, we fit the data using a Monte
Carlo Markov Chain (MCMC) method provided by the COSMOMC package
\citep{lewis2002}. The best-fit results are shown in the right-hand
panels of Fig.~\ref{fig:ESD1}. In order to illustrate the typical
uncertainties and degeneracies among the various parameters,
Figs.~\ref{fig:mcmc1} and~\ref{fig:mcmc2} show the joint constraints
on a subset of parameter pairs for satellites in the
$r_p=[0.1,0.3]\,\mpch$ bin. The best-fit value for the subhalo
mass is $\log(M_{\rm sub}/\ms) = 11.68 \pm 0.67$, which is in
excellent agreement with the average subhalo mass, 
$\langle \log{M_{\rm sub,theory}/\ms} \rangle = 11.30$, 
assigned to the satellite galaxies according to the 
model described in \S\ref{sec:model}. Due to
the limited data, however, the constraint is not particularly
tight. In particular, the 95\% confidence interval for $M_{\rm sub}$
covers the entire range $\log(M_{\rm sub}/\ms) = [9.0, 12.5]$ 
In the case of the parameters $\rho_{\rm 0,sub}$ 
and $r_{\rm s,sub}$, no meaningful constraints can
be obtained due the limited amount and quality of the data on small
scales. The results for the $r_p=[0.3,0.5]\,\mpch$ bin are similar.
The best-fit values for $M$, $r_p$ and $M_{\rm sub}$ are listed in
Table~\ref{tab:para}.   Note that the best-fit value for the host
halo mass, $M_{\rm fit}\sim 10^{13.7}\ms$, is significantly
larger than the average mass obtained directly from the SDSSGC
($10^{13.37}\ms$) or from the lensing signal around the centrals
($10^{13.32}\ms$; see Fig.~\ref{fig:central}). However, this arises
because $M_{\rm fit}$ is weighted by the number of satellite
galaxies per host. Since more massive groups host more satellites,
on average, this weighting biases the inferred host halo
high \citep[cf.][]{More2009}. Indeed, if we use the group catalogue
to compute the satellite-weighted average host halo mass, we obtain
the values listed in the fourth column of Table~\ref{tab:para}, 
which are in much better agreement with the best-fit value for
$M_{\rm fit}$.
  
In the group catalog, some galaxies that are identified as satellite 
galaxies may actually be centrals of other (mostly low-mass)
haloes along the line of sight. These galaxies, which we refer 
to as interlopers, produce contaminations to the total lensing 
signal. In Li13, the effect of interlopers was investigated 
with the mock group catalogue given in \citet{Yang2007} 
(for more details, see Sec.6.2 in L13). Using the same method, 
we find that the fraction of interlopers in the groups used here 
is 13\%. The bias produced by the interlopers in the estimated 
subhalo mass is $\sim 0.1$ dex, much smaller than the 
statistical errors.

\begin{table*}
\begin{center}
  \caption{The best-fit values of model parameters and our theoretical
    predictions.  $N_{\rm sat}$ is the number of satellites used as
    lenses. $\langle \log M_*\rangle$ is the average stellar mass of
    the satellites.  $\langle \log{M_{\rm theory}} \rangle $ is the
    average, satellite-weighted host halo mass predicted directly from
    the galaxy group catalog. $\langle \log{M_{\rm sub,theory}}
    \rangle$ is the mean subhalo mass expected from our theoretical
    model (\S~\ref{sec:model}).  $\log{M_{\rm fit}} $, $r_{\rm p,fit}$
    and $\log{M_{\rm sub,fit}}$ are the best-fit values for $M$, $r_p$
    and $M_{\rm sub}$ obtained using the model described in
    \S~\ref{sec:result}. All errors indicate the 68\% confidence
    intervals. Finally, $\chi^2_{\rm red}$ is the reduced $\chi^2$ of
    the best-fit model. Masses and distances are in units of $\ms$ and
    $\mpch$, respectively.}

\begin{tabular}{c|c|c|c|c|c|c|c|c}
 \hline   
 &&&&&&&\\
      Satellite $r_p$ range
& $N_{\rm sat}$ 
& $\langle \log(M_{*}) \rangle$ 
& $\langle \log{M_{\rm theory}} \rangle $
& $\log{M_{\rm fit}} $ 
& $r_{\rm p,fit}$ 
& $\langle \log{M_{\rm sub,theory}} \rangle $ 
& $\log{M_{\rm sub, fit}}$ 
& $\chi^2_{\rm red}$ \\

   &&&&&&&\\
  \hline 
   & & & & & &&\\
 $[0.1, 0.3]$ &  595 & 10.51& 13.67&  13.73 $\pm$ 0.04    &  0.19$\pm$ 0.01  & 11.30   & 11.68$\pm$ 0.67 & 1.68\\
   & & & & & &&\\
 \hline
  &&&&&&&\\
 $[0.3,0.5]$ & 475   &  10.52 & 13.78 & 13.50  $\pm$ 0.08 &  0.33 $\pm$ 0.04  & 11.23   &11.68$\pm$ 0.76  & 1.02 \\
  &&&&&&&\\
 \hline
\end{tabular}
\label{tab:para}
\end{center}
\end{table*}

\section{Summary}
\label{sec:sum}

We have used the  \citet{Yang2007} galaxy group catalog
constructed from the SDSS spectroscopic survey to select satellite
galaxies and obtained tangential shears around them using sources
selected from the CS82. This has resulted in a direct 
measurement of the gravitational lensing effect due to 
dark matter subhalos associated with satellite galaxies. 
Compared with previous studies based on massive clusters
of galaxies \citep[e.g.][]{Natarajan2007, Natarajan2009},
our results present the first measurement of the subhalo 
masses of satellites in galaxy groups.

The lensing effect is measured for satellites in groups with masses in
the range $[10^{13},5\times10^{14}]\ms$, and the results agree well
with theoretical expectations, although the errorbars are quite large,
especially on small scales.  Fitting the data points with a truncated
NFW profile, we obtain an average subhalo mass of 
$\log (M_{\rm sub}/\ms) = 11.68 \pm 0.67$ for satellites located at projected
group-centric distances in the range $[0.1, 0.3]\mpch$, 
and $\log(M_{\rm sub}/\ms ) = 11.68 \pm 0.78$ for those in the range
$[0.3,0.5]\mpch$. The current data is still insufficient to
put any meaningful constraints on the central density, 
$\rho_{\rm 0,sub}$, and/or scale radius, $r_{\rm s,sub}$, of subhalos.
The best-fit subhalo masses are  consistent (within the errors)
with the truncated subhalo masses assigned to satellite galaxies using
abundance matching.  Our results prove the feasibility of using
galaxy-galaxy weak lensing to study the properties of subhalos, once a
well-defined galaxy group catalog is available to  pre-select
satellite galaxies.  As discussed in L13, with next generation weak
lensing surveys, which will yield many more source galaxies behind
many more foreground galaxy groups, one will be able to
constrain both the mass and the structure of subhalos associated
with satellite galaxies in narrow bins of host halo mass bins and
group-centric distance, $r_p$. This will yield constraints on the
formation and evolution of dark matter subhalos, and perhaps even on
the nature of the dark matter through its impact on the formation of
cosmic structure on small scales.

\section*{Acknowledgements}

Based on observations obtained with MegaPrime/MegaCam, a joint project of CFHT and CEA/DAPNIA, at the Canada-France-Hawaii Telescope (CFHT), which is operated by the National Research Council (NRC) of Canada, the Institut National des Science de l'Univers of the Centre National de la Recherche Scientifique (CNRS) of France, and the University of Hawaii. The Brazilian partnership on CFHT is managed by the Laboratório Nacional de Astronomia (LNA). This work made use of the CHE cluster, managed and funded by ICRA/CBPF/MCTI, with financial support from FINEP and FAPERJ. We thank the support of the Laboratório Interinstitucional de e-Astronomia (LIneA). We thank the CFHTLenS team for their pipeline development and verification upon which much of this surveys pipeline was built.

LR acknowledges support from NSFC, grant NO.11303033.
HYS acknowledges the support from Marie-Curie International
 Incoming Fellowship (FP7-PEOPLE-2012-IIF/327561), 
 Swiss National Science Foundation (SNSF) and NSFC of China under grants 11103011.
HJM acknowledges support of NSF AST-0908334
and NSF AST-1109354.
JPK acknowledges support from the ERC advanced grant LIDA and from CNRS.
XHY acknowledges support from NSFC (Nos. 11128306, 11121062, 11233005).
TE is supported by the Deutsche
Forschungsgemeinschaft through project ER 327/3-1 and by the Transregional
Collaborative Research Centre TR 33 - "The Dark Universe".

\bibliography{lensing}

\end{document}